\documentclass[aps,pr,twocolumn,groupedaddress,notitlepage,
floatfix,superscriptaddress]{revtex4-1}

\pdfoutput=1
\usepackage{graphicx,graphics,epsfig,subfigure,times,bm,bbm,amssymb,amsmath,amsthm,mathrsfs,MnSymbol}
\usepackage{gensymb}
\usepackage{amsfonts}
\usepackage{float}
\usepackage[matrix,frame,arrow]{xypic}
\usepackage[pdfstartview=FitH]{hyperref}

\usepackage[pdftex]{color}
\usepackage{braket}  
\usepackage{enumerate}
\usepackage[normalem]{ulem}
\usepackage[usenames,dvipsnames]{xcolor}
\usepackage{multirow}
\usepackage{mathtools}
\usepackage{bbm}
\usepackage{titletoc}

\definecolor{orange}{rgb}{1,0.5,0}

\newcommand{\RNum}[1]{\uppercase\expandafter{\romannumeral #1\relax}}

\newcommand{\ignore}[1]{}
\usepackage{geometry}

\ignore{
	\documentclass[eprintnumbers,amsmath,amssymb,onecolumn,a4paper,caption ]{article}
	\usepackage{amsfonts}
	\usepackage{amssymb}
	\usepackage{mathrsfs}
	\usepackage{mathbbold}
	\usepackage{bbm}
	\usepackage{mathrsfs}
	\usepackage{dcolumn}
	\usepackage{bm}
	\usepackage{times,epsfig,amssymb,amsmath}
	\usepackage{float}
	\usepackage{subfigure}
	\usepackage{geometry}\geometry{left=2.5cm,right=2.5cm,top=3cm,bottom=3cm}

	\usepackage{color}

}

\hypersetup{
	colorlinks=true,       
	linkcolor=red,          
	citecolor=blue,        
	filecolor=magenta,      
	urlcolor=blue,           
	runcolor=cyan
}

\begin{document}
	\title{Quantum Simulation of Lattice Gauge Theories on Superconducting Circuits: Quantum Phase Transition and Quench Dynamics}
	
	\author{Zi-Yong~Ge 
	}
	\affiliation{Beijing National Laboratory for Condensed Matter Physics, Institute of Physics, Chinese Academy of Sciences, Beijing 100190, China}
	\affiliation{School of Physical Sciences, University of Chinese Academy of Sciences, Beijing 100190, China}
	
	\author{Rui-Zhen Huang 
	}
	\email{huangruizhen13@mails.ucas.ac.cn}
	\affiliation{Kavli Institute for Theoretical Sciences, University of Chinese Academy of Sciences, Beijing 100190, China}

	\author{Zi Yang~Meng 
	}
	\affiliation{Beijing National Laboratory for Condensed Matter Physics, Institute of Physics, Chinese Academy of Sciences, Beijing 100190, China}
	\affiliation{Songshan Lake Materials Laboratory, Dongguan 523808, China}
	\affiliation{Department of Physics and HKU-UCAS Joint Institute of Theoretical and Computational Physics, The University of Hong Kong, Pokfulam Road, Hong Kong SAR, China}

	\author{Heng~Fan 
	}
	\email{hfan@iphy.ac.cn}
	\affiliation{Beijing National Laboratory for Condensed Matter Physics, Institute of Physics, Chinese Academy of Sciences, Beijing 100190, China}
	\affiliation{School of Physical Sciences, University of Chinese Academy of Sciences, Beijing 100190, China}
	\affiliation{Songshan Lake Materials Laboratory, Dongguan 523808, China}
	\affiliation{CAS Center for Excellence in Topological Quantum Computation, UCAS, Beijing 100190, China}

\begin{abstract}
	Recently, quantum simulation of low-dimensional  lattice gauge theories (LGTs)
	has attracted many interests, which may improve our understanding of strongly correlated quantum many-body systems.
	Here, we propose an implementation to approximate $\mathbb{Z}_2$ LGT on superconducting quantum circuits, where the effective theory is a mixture of a LGT and a gauge-broken term. Using matrix product state based methods, both the ground state properties and quench dynamics are systematically investigated. With an increase of the transverse (electric) field, the system displays a quantum phase transition from a disordered phase to a translational symmetry breaking phase. In the ordered phase, an approximate Gauss law of the $\mathbb{Z}_2$ LGT emerges in the ground state. Moreover, to shed light on the experiments, we also study the quench dynamics, where there is a dynamical signature of the spontaneous translational symmetry breaking. The spreading of the single particle of matter degree is diffusive under the weak transverse field, while it is ballistic with small velocity for the strong field. Furthermore, due to the emergent Gauss law under the strong transverse field, the matter degree can also exhibit confinement dynamics which leads to a strong suppression of the nearest-neighbor hopping.
	Our results pave the way for simulating the LGT on superconducting circuits, including the quantum phase transition and quench dynamics.
\end{abstract}

\maketitle

\section{Introduction}
In the last two decades, with the rapid development of quantum manipulation technologies, simulating quantum physics  on synthetic quantum many-body systems, termed quantum simulation~\cite{Buluta2009,Georgescu2014}, has attracted enormous interests.
Quantum simulations are expected to provide an alternative for solving problems in strongly correlated systems, which may be much challenging by traditional methods.
Meanwhile, quantum simulations have also greatly enriched the study of quantum physics,  especially the out-of-equilibrium dynamics of quantum many-body systems~\cite{Polkovnikov2011,Eisert2015}.
There are various platforms for performing  quantum simulations, such as ultracold atoms, trapped ions, nuclear spins, superconducting circuits, and so on.
Among them, superconducting circuits~\cite{Makhlin2001,Gu2017} are believed to be one of the most competitive candidates for achieving the universal quantum computation, due to the scalability, long coherent time, and high-precision control. Specifically, each superconducting qubit can be precisely and simultaneously addressed  under arbitrary bases. Thus, it is believed that the superconducting circuit platform is another good choice to perform quantum simulations of challenging quantum many-body physics, and recently, there are many experiments about simulating the dynamics of quantum many-body systems in superconducting circuits~\cite{Xu2018,Roushan2017,Salathe2015,Barends2015,Zhong,Song1,Flurin2017,Ma2019,Yan2019,Ye2019,Guo2019,Xu2020}.
Furthermore, the quantum supremacy~\cite{Arute2019,Zhong2020,Wu2021} has been demonstrated on a superconducting processor.

Lattice gauge theories (LGTs)  are originally proposed to understand the confinement of quarks~\cite{Wilson1974}.
In addition to high energy physics, LGTs also play a significant role in condensed matter physics~\cite{Kogut1979,Wen2004,Fradkin2013,Kitaev2003_1,Kitaev2003_2}.
In low-dimensional condensed matter systems, gauge fields can emerge due to the strong quantum fluctuation. For instance, the emergent gauge fields are closely related to the quantum spin liquid~\cite{Zhou2016}, deconfined  quantum critical point~\cite{Senthil2004}, and so on.
Recently, the real-time dynamics~\cite{Hebenstreit2013,Kormos2017} and quantum simulations of LGTs and  have drawn much attention from both theoretical and experimental physicists.
The corresponding schemes have been proposed in terms of ultracold atoms~\cite{Zohar2012,Banerjee2012,Barbiero2019}, trapped ions~\cite{Hauke2017}, specific superconducting circuits~\cite{Marcos2013,Brennen2016}, and other digital/analogue quantum circuits~\cite{Zohar2017,Chamon2020}. Some of these schemes have been realized in ultracold atoms, experimentally~\cite{Schweizer2019,Yang2019,Gorg2019}.
Nevertheless, it is still a challenge to study LGT on superconducting circuits, experimentally, and the key problem is how to design a feasible scheme for realizing a LGT Hamiltonian.

In this work, we focus on one-dimensional (1D) superconducting quantum circuits~\cite{Makhlin2001,Gu2017}.
By applying stagger longitudinal ($z$-directional) and transverse ($x$-directional) fields, we obtain an effective Hamiltonian containing a $\mathbb{Z}_2$ LGT and a gauge-broken term.
The transverse field can induce a quantum phase transition (QPT) from a gapless disorder phase to a translation-symmetry breaking the dimer phase.
Despite  lacking $\mathbb{Z}_2$ gauge structure for the whole Hamiltonian, an approximate Gauss law can still emerge in the ground state of dimer phase.
Furthermore, inspired by the experiment, we also study the real-time dynamics, during which there is a dynamical signature of translational symmetry breaking.
We find that the particle of matter degree can diffuse to the whole system under weak transverse fields after a quantum quench.
However, in the case of strong transverse fields, it displays a ballistic propagation.
Moreover, the nearest-neighbor (NN) hopping almost disappears, which is a strong dynamical signature of confinement induced by the emergent approximate Gauss law.

\section{The model}
\begin{figure}[t]     		
	\includegraphics[width=0.48\textwidth]{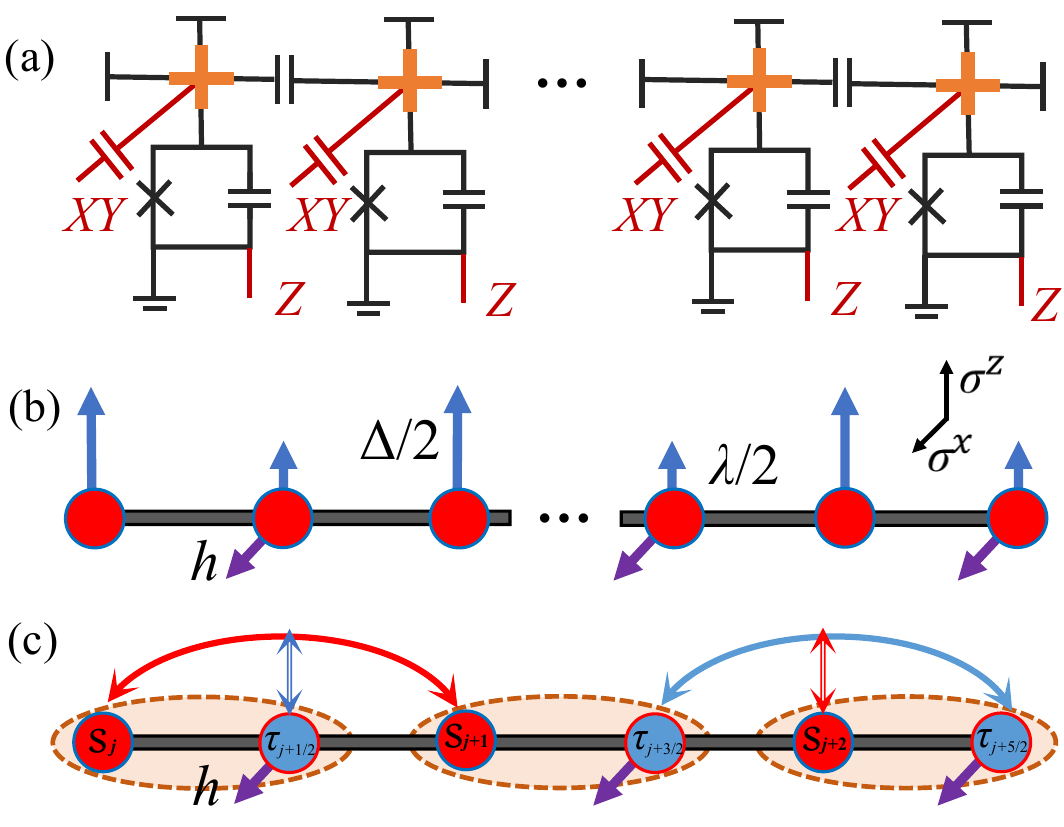}
	\caption{(a) The diagram of 1D superconducting quantum circuits. The nearest-neighbor two qubits are coupled capacitively, and each qubit contains a Z bias line and an XY driven line, which can be used to control the frequency and transverse field, respectively.		
		(b) Lattice diagram of Eq.~(\ref{H}). The strengths of longitudinal fields are $\Delta/2$ and $\lambda/2$ for the odd and even sites, respectively.
		The transverse field is only applied on the even sites with the strength $h$.
		(c) Lattice diagram of Hamiltonian~(\ref{Hf}). The odd and even qubits are labeled by $s$ (red sites) and $\tau$ (blue sites), respectively. The arrow represent the three-body coupling, and  $s_j$ and $\tau_{j+\frac{1}{2}}$  compose an unit cell (dashed orange circles).}
	\label{fig_1}
\end{figure}

Here we consider a superconducting quantum processor containing $2L$ qubits arranged into a chain, see  Fig.~\ref{fig_1}(a),
where the Hamiltonian of this system is the 1D Bose-Hubbard model~\cite{Roushan2017,Yan2019,Ye2019}.
The on-site interaction of this Bose-Hubbard model originates from anharmonicity of each qubit, which is much larger than the NN coupling strength~\cite{Roushan2017,Yan2019,Ye2019},  so the Fock space of each qubit can be truncated to 2.
Thus, the system can be described by an isotropic 1D $XY$ model,
of which the Hamiltonian reads~\cite{Roushan2017,Yan2019,Ye2019}
\begin{align} \label{Hxy}
	\hat H =\sum_{j}^{2L-1}J_{j}(\hat{\sigma}^+_{j} \hat{\sigma}^-_{j+1}+\text{H.c.})+\sum_{j=1}^{2L} (\frac{V_{ j}}{2}\hat{\sigma}^z_{j}
	+ h_{ j}\hat{\sigma}_{j}^x),
\end{align}
where $\hat{\sigma}^{\pm} = (\hat{\sigma}^x\pm i\hat{\sigma}^y)/2$, $\hat{\sigma}^{x,y,z}$
are Pauli matrices, $J_{j}$ is the NN coupling strength,
$V_j$ is the local potential which can be tuned by Z pulse, and $h_j$ is the transverse field which can be controlled by XY driven.
For simplicity, we consider a homogeneous coupling strength in the following discussion, i.e., $J_j=J$.

To implement $\mathbb{Z}_2$ LGT,
we let the local potentials at odd and even qubits are $\Delta$ and $\lambda$, respectively, and let detuning between two NN qubits much larger than the coupling strength, i.e., $|\Delta-\lambda|\gg J$.
In addition, the transverse field is only applied on the even qubits with equal strength $h$, see Fig.~\ref{fig_1}(b).
Therefore, the Hamiltonian (\ref{Hxy}) can be simplified as
\begin{align} \label{H} \nonumber
	\hat H =\sum_{j}^{2L-1}&J(\hat{\sigma}^+_{j} \hat{\sigma}^-_{j+1}+\hat{\sigma}^+_{j+1} \hat{\sigma}^-_{j})+\sum_{j=1}^{L} (\frac{\Delta}{2}\hat{\sigma}^z_{2j-1}\\
	+& \frac{\lambda}{2}\hat{\sigma}^z_{2j}+  h\hat{\sigma}_{2j}^x),
\end{align}
where $\Delta\gg J,\lambda, h$.

\begin{figure*}[tb]     		
	\includegraphics[width=0.95\textwidth]{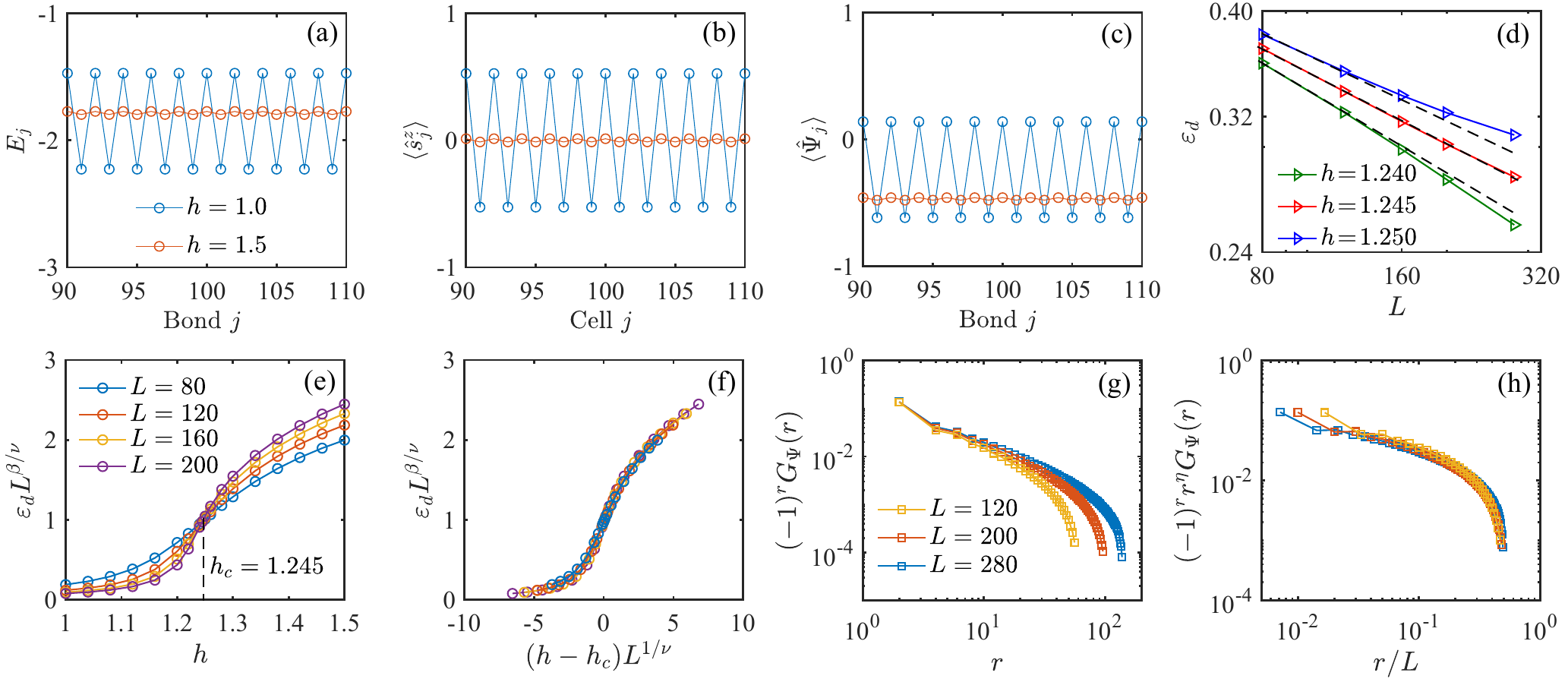}
	\caption{The distribution of (a) $E_j$, (b) $\braket{\hat s^z_j}$, and (c) $\braket{\hat \Psi_j}$ in the ground state with $L=200$. (d) The function between $\varepsilon_d$ and system size $L$. At the critical point $h=h_c=1.245$, $\varepsilon_d\sim L^{-\Delta_\varepsilon}\sim L^{-0.23}$. The dashed black lines are the linear fittings. (e) Rescaled $\varepsilon_dL^{\Delta_\varepsilon}$ as the function of $h$ for different system sizes.
		The curves are across at the critical point. (f) The data collapse of (e) with $1/\nu\approx0.62$. (g) The correlation functions $G_\Psi (L/2-r,L/2)$ for different system sizes at the critical point. (h) The data collapse of (g) with $\eta\approx0.46$.
	}	
	\label{fig_2}
\end{figure*}

Using Schrieffer-Wolff transformation~\cite{Schrieffer1966,Bravyi2011}, we can obtain an effective Hamiltonian $\hat H_f$ written as
\begin{align} \label{Hf} \nonumber
	& \hat H_f =\hat H_1 + \hat H_2, \\ \nonumber
	&\hat  H_1 =\sum_{j=1}^{L-1}g_e(\hat{s}^+_{j} \hat{\tau}^z_{j+\frac{1}{2}}\hat{s}^-_{j+1} +
	\text{H.c.} ) +\sum_{j=1}^{L}h\hat{\tau}_{j+\frac{1}{2}}^x ,\\
	& \hat H_2 = -\sum_{j=2}^{L}g_e(\hat{\tau}^+_{j-\frac{1}{2}} \hat{s}^z_{j}\hat{\tau}^-_{j+\frac{1}{2}}+ \text{H.c.}),
\end{align}
where $\hat s_j:=\hat\sigma_{2j-1}$ and $\hat\tau_{j+\frac{1}{2}}:=\hat\sigma_{2j}$ label the odd and even qubits, respectively, and $g_e \approx J^2/\Delta$ is the effective three-body coupling strength, see Fig.~\ref{fig_1}(c).
The detailed derivation of  effective Hamiltonian is presented in Supplementary Materials.
Here, $\hat H_1$ is nothing but a typical $\mathbb{Z}_2$ LGT coupled with a matter field~\cite{Schweizer2019,Borla2020}, where $\tau$ and $s$  are gauge and matter fields, respectively, and the transverse field $h$ is the corresponding $\mathbb{Z}_2$ electric field.
The  $\mathbb{Z}_2$ gauge transformation can be defined as $\hat{Q}_j = \hat{\tau}_{j-\frac{1}{2}}^x\hat{s}_j^z\hat{\tau}_{j+\frac{1}{2}}^x$ satisfying $[\hat Q_j, \hat H_1]=0$.
However, $[\hat Q_j, \hat H_2]\neq0$, so that $\hat H_f$  lacks $\mathbb{Z}_2$ gauge invariance.
In the following discussion, we will show that an approximate $\mathbb{Z}_2$ Gauss law can emerge in the ground state.
In addition, the $s$ sector has a global $U(1)$ symmetry, so that the total $s$ spins $[\sum_{j=1}^L \hat{s}_j^z,\hat H_f]=0$ are conserved,
and $\hat H_f$ is translation invariant.

\section{Ground-state phase diagram}
We use density matrix renomalized group (DMRG) method~\cite{Schollwock2005,Schollwock2011} to study the ground-state properties of $\hat H_f$.
Here, we fix the coupling $g_e = 1$,  set $h$ as the driving parameter, and only consider $h>0$. Open boundary condition is used for the numerical simulation. Furthermore, we set the $s$ spins to be half-filling, i.e., $\sum_{j=1}^L \hat{s}_j^z=0$.
In the numerical calculation, we choose the max bond dimension up to 600 with truncation error smaller than $10^{-7}$. Furthermore,
we adopt careful finite truncation error analysis to obtain more accurate results.
The details of our numerical method are shown in Supplementary Materials.

When $h=0$, $\hat{H}_f$ describes the free gapless fermions after the Jordan-Wigner transformation. We expect the system belongs to Luttinger liquid at small $h$. The transverse field term serves as a relevant perturbation and may induce a gap. Indeed, from the numerical simulation, we find that there is a QPT and the translational symmetry is spontaneously broken at large $h$.
Here, to probe translational symmetry broken, we calculate the local energy density $E_j :=\langle { \hat H_j} \rangle $, where
\begin{align} \label{Hj} \nonumber
	\hat  H_j =&g_e (\hat{s}^+_{j} \hat{\tau}^z_{j+\frac{1}{2}}\hat{s}^-_{j+1} -\hat{\tau}^+_{j+\frac{1}{2}} \hat{s}^z_{j}\hat{\tau}^-_{j+\frac{3}{2}}
	+\text{H.c.} ) \\
	&+\frac{h}{2}(\hat{\tau}_{j+\frac{1}{2}}^x+\hat{\tau}_{j+\frac{3}{2}}^x) ,
\end{align}
and  $\langle \cdot\rangle$ represents taking expectation value towards the ground state.
As shown in Fig.~\ref{fig_2}(a),
the local energy density in the bulk retains translational invariance for small $h$,  while it becomes dimerized in the large $h$ region.
It can also be reflected by the polarization of the $\hat s^z$, see Fig.~\ref{fig_2}(b). This reveals the condensation of $\hat{\tau}^x$ induces a $\pi$ momentum which breaks the translational invariance.

Next, we use finite-size scaling to determine the critical properties.
From Fig.~\ref{fig_2}(c), we can find that $\hat \Psi_{j} := \hat{\tau}^+_{j-\frac{1}{2}} \hat{s}^z_{j}\hat{\tau}^-_{j+\frac{1}{2}}+ \text{H.c.}$  is the other term breaking the translation invariance.
Thus, we choose the order parameter as the expectation value difference of the operator $\hat \Psi_{j}$ between odd and even bonds.
To avoid the boundary effect, we only consider the central two bonds, i.e.,  the order parameter $\varepsilon_d:=\braket{\hat \Psi_{L/2}-\hat \Psi_{L/2-1}}$.
At the critical point $h=h_c$, $\varepsilon_d$ satisfies a power law with respect to the system size $L$, i.e.,
$\varepsilon_d|_{h=h_c} \sim L^{-\Delta_\varepsilon}$. According to Fig.~\ref{fig_2}(d), we determine the critical point to be $h_c=1.245$ and $\Delta_\varepsilon \approx0.23$. To double check the accuracy of our calculation, we construct the dimensionless quantity $\varepsilon_dL^{\Delta_\varepsilon}$ and find it indeed crosses at $h_c$ for different sizes as shown in Fig.~\ref{fig_2}(e). Then from the data collapse of the order parameter we obtain the exponent $1/\nu\approx0.62$, see Fig.~\ref{fig_2}(f). We also study the correlation $ G_\Psi (j-i) = \braket{\hat \Psi_{j}\hat \Psi_{i}}-\braket{\hat \Psi_{j}}\braket{\hat \Psi_{i} }$
($\hat{\Psi}_i$ has the same dimension as $\hat{\Psi}_{2j}-\hat{\Psi}_{2j-1}$.).
As shown in Figs.~\ref{fig_2}(g,h), $G_\Psi (r)$ exhibits a power-law decay at $h_c$ with the exponent $\eta\approx0.46$ which is consistent with $\eta = 2 \Delta_\varepsilon$.
In Supplementary Materials, we present the additional numerical results of correlation function away from the critical point.

\begin{figure}[tb]     		
	\includegraphics[width=0.48\textwidth]{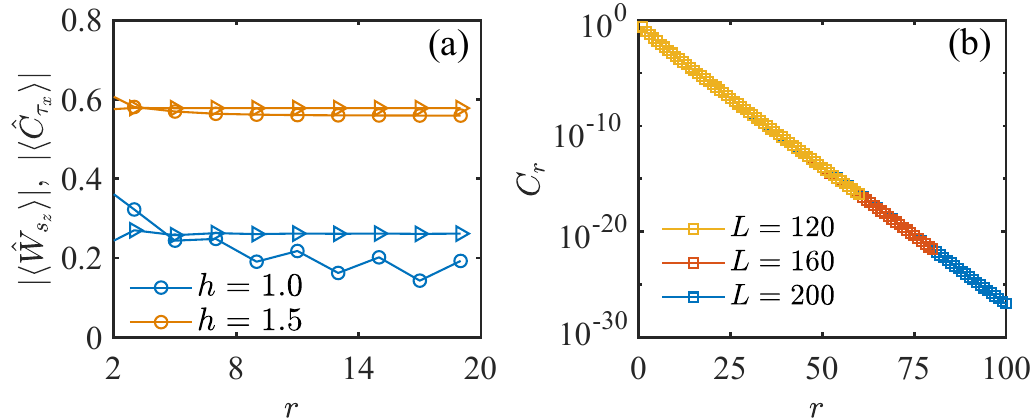}
	\caption{(a) The expectation values of $|\braket{\hat W_{s_z}(r)}|$ (circle) and $|\braket{\hat C_{\tau_x}(r)}|$ (triangle) in the ground state with $L=80$ for $h=1$ and $h=1.5$, respectively. To avoid the boundary effect, we calculate  these two objects at the cell interval $[ L/2-r/2+1$, $L/2+r/2]$. (b) The gauge invariant correlation function $G_{L/2,L/2+r}$. }
	\label{fig_3}
\end{figure}

Here, we explore the fate of $\mathbb{Z}_2$ Gauss law in the ground state.
We define the $\mathbb{Z}_2$ charge $\hat W_{s_z} (i,j):= \prod_{i\leq k\leq j} \hat s_k^z$ and the flux $\hat C_{\tau_x}(i,j) := \hat \tau_{i-\frac{1}{2}}^x\hat \tau_{j+\frac{1}{2}}^x$.
We know that $\hat W_{s_z} (i,j)= \pm\hat C_{\tau_x}(i,j)$ must be  satisfied rigorously for  $\mathbb{Z}_2$ gauge invariant systems (e.g., $\hat H_1$), which is Gauss law of $1D$ $\mathbb{Z}_2$ LGT.
Surprisingly from Fig.~\ref{fig_3}(a), one can find, in our system, $\langle {\hat C_{\tau_x}(i,j)}\rangle\simeq (-1)^{f(i,j)}\langle {\hat W_{s_z} (i,j)}\rangle$ for large $h$.
Here, $f(i,j)=\frac{(j-i+1)(j+i+2)}{2}$ is an integer, and
the sign of charge $\hat W_{s_z}$ is consistence with the polarization of $\hat s_z$.
However,  there is a large deviation between $\langle {\hat C_{\tau_x}(i,j)}$ and $\langle {\hat W_{s_z} (i,j)}\rangle$ for small $h$. 
This means that an approximate $\mathbb{Z}_2$ Gauss law emerges in the ground state of $\hat H_f$ for the strong transverse field, although the $\mathbb{Z}_2$ gauge invariance is absent in $\hat H_f$. Therefore the dimer state can be regarded as an analogy of the confined phase of the $\mathbb{Z}_2$ LGT.
To further illustrate the confined phase, we calculate the gauge invariant correlation function defined as $G_{i,j} = \braket{\hat s_i^+ (\prod_{i\leq k< j} \hat{\tau}_k^z)\hat s_j^-}$~\cite{Borla2020}. Fig.~\ref{fig_3}(b) shows that $G_{i,j}$ decays exponentially, which is another evidence that $s$ sector is confined for large $h$~\cite{Borla2020}.

\section{Quench dynamics}
In quantum-simulation experiments, preparing the ground state of $\hat H_f$ is much challenging. Instead,
it is more convenient to study the quench dynamics.
Therefore, we expect that the above ground-state properties can be probed by the dynamical behaviors, which may be realized in future experiments.
Here, we use  time evolving block decimation (TEBD) method~\cite{Schollwock2005,Schollwock2011,Vidal2004} under the open boundary condition to study the real-time dynamics of $\hat H_f$. Here, we choose second-order Suzuki-Trotter decomposition and  the max bond dimension up to 400. Furthermore,
we also enlarge the maximum bond dimension and decrease time of single step till the final results converge.
The details of our TEBD methods are presented in Supplementary Materials.
We consider the dynamics in the $s$ sector during the quench process and set the total $s$ charge to $\sum_{j=1}^L \hat{s}_j^+ \hat s_j^- = 1$. The initial state is chosen as $|\psi_0\rangle = |s\rangle\otimes|\tau\rangle$, of which $|s\rangle = \ket{...\downarrow\downarrow\uparrow\downarrow\downarrow...}$ and $|\tau\rangle=|...\leftarrow\leftarrow\leftarrow...\rangle$.
That is, for $s$ sector, only the central cell is spin-up, and all $\tau$ sites are at the ground state of $\hat \tau^x$, i.e., $\ket{\leftarrow}:=\frac{1}{\sqrt{2}}(\ket{\uparrow}-\ket{\downarrow})$.
We calculate the time evolution of the density distributions of matter field $\rho_j(t) := \bra{\psi(t)} \hat{s}^+_j\hat{s}^-_j\ket{\psi(t)} $, where $\ket{\psi(t)} = \mathrm{\exp}(-i\hat{H}_f t) \vert \psi_0 \rangle$.

\begin{figure}[tb]     		
	\includegraphics[width=0.48\textwidth]{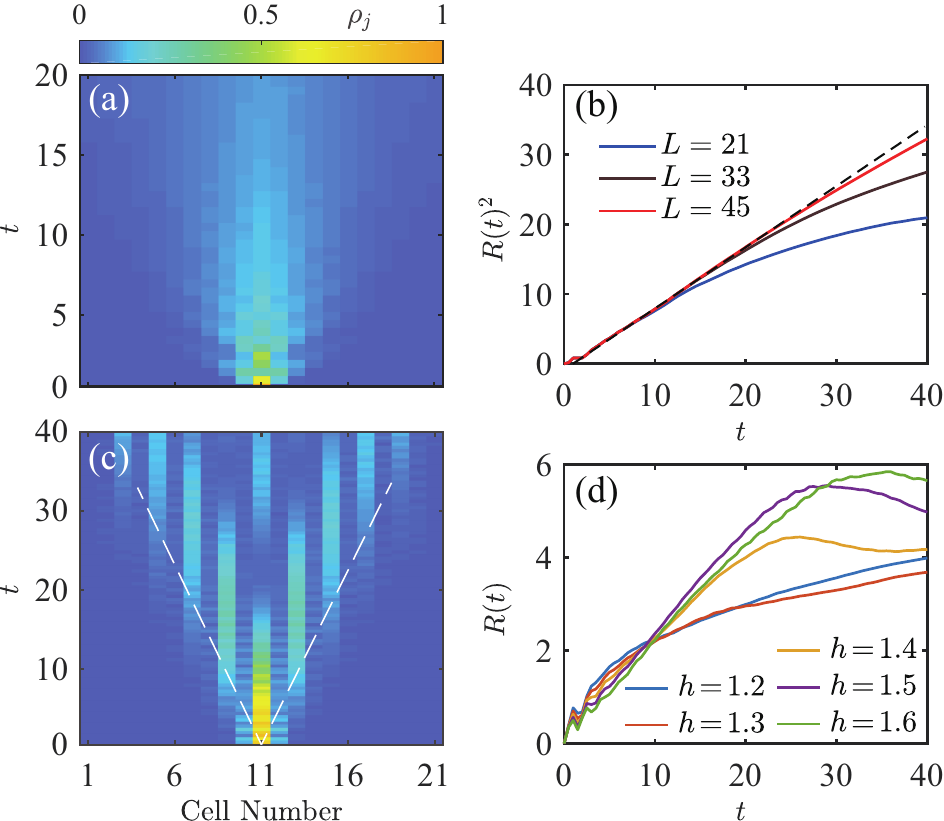}
	\caption{(a) The time evolution of $\rho_j$ for $h=1$ and $L=21$. (b) The curves of $R(t)^2$ for $h=1$ with different system sizes. The black dashed line is a linear fitting. (c) The time evolution of $\rho_j$ for $h=2$ and $L=21$. (d) The curves of $R(t)$ for different transverse field strength with $L=21$. There is a sharp transition around $h=1.3$.
	}
	\label{fig_4}
\end{figure}

We firstly study the case of small transverse fields. We find that the $s$ charge can spread to the whole system after a quantum quench as shown in Fig.~\ref{fig_4}(a). To pursue how the $s$ charge spreads, we calculate $R(t) = \sum_j^L|j-\frac{L+1}{2}|\rho_j(t)$~\cite{Kim2013}, which is the average distance to the initial position for the $s$ charge. As demonstrated in Fig.~\ref{fig_4}(b), one can find that $R(t) \propto \sqrt{t-t_0}$ in the intermediate-time regime, where $t_0$ is a time shift resulted from the relaxation at the beginning. Thus, the spreading of the $s$ charge is diffusive after a quick relaxation under the weak transverse field.
In fact, the initial state, in this case, is far away from the ground state, which makes the system thermalize.
Therefore,  $\tau$ sector can be considered as a random scatting potential of $s$ sector after the relaxation leading to a diffusive spreading of $s$ charge.

For the large transverse field, the propagation of the $s$ charge is very different.
When $h=2$, ss shown in Fig.~\ref{fig_4}(c), one can find that the $s$ charge exhibits a light-cone-like spreading at long time rather than diffusion. 
In the case of a strong transverse field, the system locates a low-energy regime during the quench dynamics, so that the system fails to thermalization.
Therefore, the $s$ charge cannot exhibit a diffusive transport.
The detailed phenomenological  discussion about these two distinct spreading types between small and large transverse fields can be found in Supplementary Materials.
In addition, we can also find that the corresponding propagation velocity is much smaller than that for the $h=0$ case (In this case, the effective Hamiltonian is $XY$ model, so the velocity is about $2g_e$~\cite{Yan2019}.). 
Meanwhile, the hopping of the $s$ spin between NN cells is strongly suppressed,
although there only exist NN hopping terms of $s$ spins  in the effective Hamiltonian $\hat H_f$.
Instead, the $s$ spin almost exhibits a next-nearest-neighbor (NNN) hopping.
In the following discussion, we will demonstrate that this suppression of  NN propagation $s$ spin  is closely related confinement induced by emergent gauge invariance.
Furthermore, we also present the curves of $R(t)$ for different $h$ in Fig.~\ref{fig_4}(d). One can find there is a clear transition from diffusive to ballistic spreading around $h=1.3$ which is close to $h_c$. Therefore, the spreading of $s$ provides a dynamical signature of the spontaneous translational symmetry breaking.

\begin{figure}[tb]     		
	\includegraphics[width=0.48\textwidth]{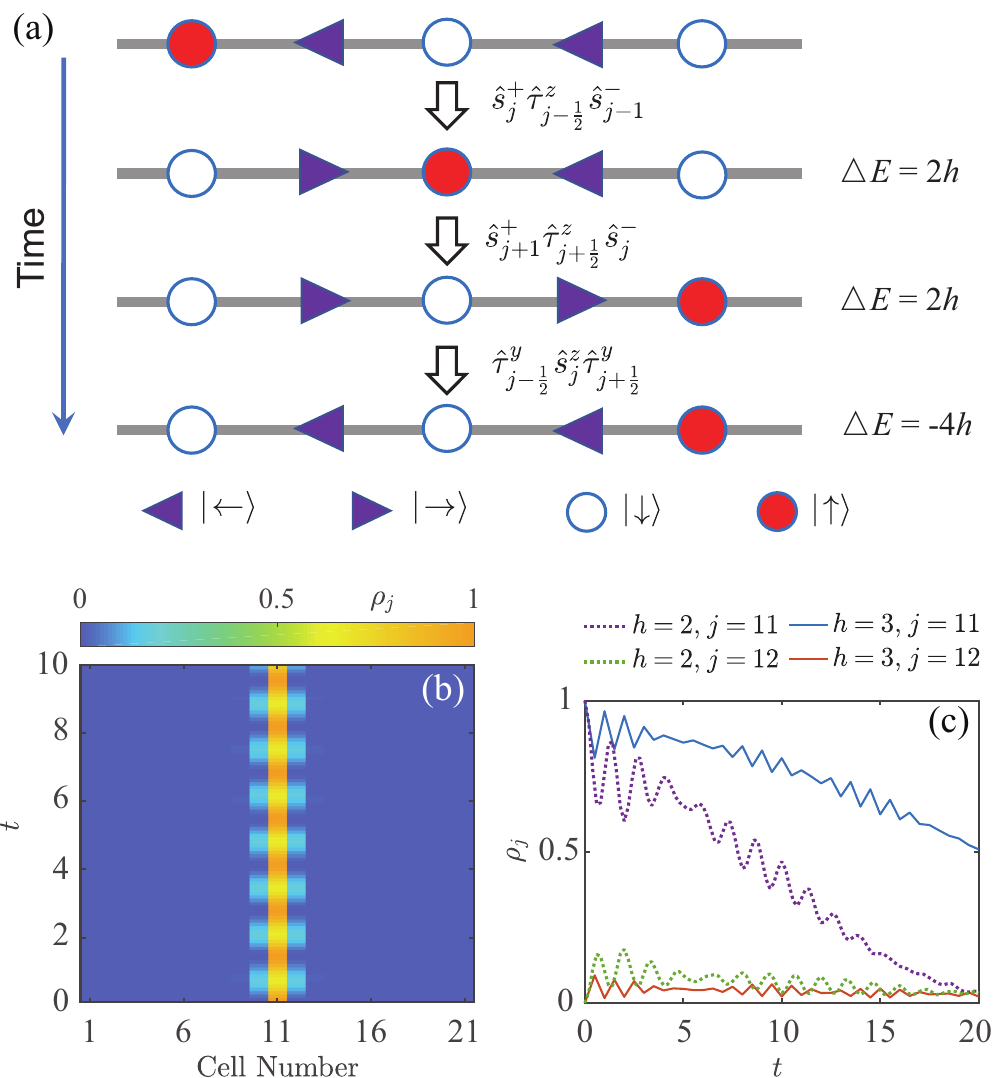}
	\caption{(a) Schematic diagram of the propagating of $s$ spin.
		(b) The time evolution of $\rho_j$ for $\hat H_1$ with $h=2$, which is consistent with the result in Ref.~\cite{Schweizer2019}.
		(c) The curves of $\rho_j(t)$ for $\hat H_f$ dynamics with $h=2$ and $h=3$. The initial state is the same as one in Fig.~\ref{fig_4}.}
	\label{fig_5}
\end{figure}

We further pursue the relation between the emergent approximate Gauss law and the anomalous spreading of $s$ charges after a quantum quench.
Clearly, without the gauge breaking term $\hat{H}_2$, the hopping of $s$ charge is accompanied by the flip of $\tau$ spin between $\ket{\rightarrow}$ and $\ket{\leftarrow}$ leading to the energy change of $2h$, see the first and second steps in Fig.~\ref{fig_5}(a).
Thus, $s$ charge is confined and localized under non-zero transverse fields~\cite{Schweizer2019,Borla2020}, as shown in Fig.~\ref{fig_5}(b).

However, what is the fate of this confinement when adding a gauge-violation term?
Here, $\hat H_2$ contains the gauge violation term $\hat{\tau}_{j-\frac{1}{2}}^y\hat{s}_j^z\hat{\tau}_{j+\frac{1}{2}}^y$.
This term can flip two NN $\tau$ spins  between $\ket{\rightarrow}$ and $\ket{\leftarrow}$.
Thus, as shown in Fig.~\ref{fig_5}(a), under the role of $\hat H_1$, the system can firstly have two imaginary processes to make a $s$ spin hop to the NNN cell by flipping two $\tau$ spins.
Then, these two $\tau$ spins can flip back through the term $\hat{\tau}_{j-\frac{1}{2}}^y\hat{s}_j^z\hat{\tau}_{j+\frac{1}{2}}^y$.
Here,  this whole process does not violate  energy conservation and thus is a real process.
This is the phenomenological mechanism why $s$ charge can only emerge NNN hopping for large $h$,  and the compression of NN hopping of $s$ charge can be considered as the effect of confinement induced by emergent Gauss law. From Fig.~\ref{fig_4}(c) and Fig.~\ref{fig_5}(b), we can find that, in the early time scale $t\le6$, the dynamics of $H_f$ is almost same as the pure gauge-invariant model, $\hat H_1$. At a much larger time scale $t>10$, higher-order process that violates the emergent gauge invariance contributes, and the NNN spreading dynamics recovers.

To further support the analysis, we also present the density distribution near the center unit cell for a larger $h$ in Fig.~\ref{fig_5}(c). Indeed the spreading becomes much weaker for a larger $h$, in which the confinement effect is stronger. Note that in the experiments, the initial time stage is the most accessible, therefore, one can observe strong signals for the confined (localized) dynamics.  These results demonstrate the emergent approximate Gauss law and confinement can indeed be witnessed by the quench dynamics.

\section{Summary}
We have constructed an effective model on the superconducting circuits, which is the mixture of a $\mathbb{Z}_2$ LGT and a gauge-broken term, and systematically studied the QPT and quench dynamics. We find that this system can exhibit a QPT from a  disorder phase to a dimer phase with the increase of the transverse field. Moreover, an approximate Gauss law can emerge for the large $h$ regime, and the dimer state can be regarded as a confinement phase as an analogy to the LGT. This QPT can also be probed by the spreading of the matter particle during the quench dynamics. It displays a transition from diffusive to ballistic spreading with an increase of the transverse field.
Remarkably, the quench dynamics for the large $h$ can be explained from the confinement viewpoint. Our work may inspire the future study of LGTs from the dynamical viewpoint. Meanwhile, our results lay the foundation for the further quantum simulation of 2D LGTs related problems on superconducting circuits.

In Refs.~\cite{Marcos2013,Brennen2016}, two schemes of implementing 1D and 2D quantum link models  on the specific superconducting circuits are proposed, respectively.
Comparing with these two works, our scheme bases on the conventional 1D superconducting circuits and manipulation technologies and is more accessible in experiments based on the state-of-the-art experimental technologies~\cite{Xu2018,Roushan2017,Salathe2015,Barends2015,Zhong,Song1,Flurin2017,Ma2019,Yan2019,Ye2019,Guo2019,Xu2020}.
On the other hand, the realization of the effective model in this work is distinct from that in cold atoms, of which gauge fields are realized by  the density-dependent tunneling matrix. While in our proposal, the matter field ($\tau$)  and gauge field ($s$)  are directly realized by the odd and even qubits, respectively.
Finally, there still exist many other open questions.
For instance, the microscopic mechanism of the emergent gauge invariance is still not clear, and it deserves further study via careful bosonization like techniques.
The thermalization of this system is also believed to be an interesting topic, and another meaningful issue is that whether we can realize a truly gauge-invariant Hamiltonian on superconducting circuits.

\begin{acknowledgements}
	The DMRG and TEBD calculations are carried out with TeNPy Library~\cite{Hauschild2018}.
	R.Z.H is supported by China Postdoctoral Science Foundation (Grant No. 2020T130643), the Fundamental Research Funds for the Central Universities and the National Natural Science Foundation of China (Grants No. 12047554).
	Z. Y. M acknowledges support from the National Key Research and Development Program of China (Grant No. 2016YFA0300502) and the Research Grants Council of Hong Kong SAR China (Grant No. 17303019).
	H. F acknowledges support from the
	National Key R \& D Program of China (Grant Nos. 2016YFA0302104 and 2016YFA0300600), National
	Natural Science Foundation of China (Grant Nos. 11774406 and 11934018), Strategic
	Priority Research Program of Chinese Academy of Sciences (Grant No. XDB28000000),
	Beijing Academy of Quantum Information Science (Grant No. Y18G07).\\
\end{acknowledgements}

\appendix

\section{Derivation of the Effective Hamiltonian}\label{App1}
In this section, we present more details of derivations from Eq.~\ref{H} to Eq.~\ref{Hf} in the main text.
As mentioned in the main text, the original Hamiltonian of the system can be written as
  \begin{align} \label{Hh} \nonumber
&\hat H = \hat H_0 + \hat H_I,\\ \nonumber
&\hat H_0= \sum_{j=1}^{L} \Delta\hat{\sigma}^+_{2j-1}\hat{\sigma}^-_{2j-1}
+\lambda\hat{\sigma}^+_{2j}\hat{\sigma}^-_{2j},\\
&\hat H_I = J\sum_{j}^{L}\hat{\sigma}^+_{2j} \hat{\sigma}^-_{2j+1}+\hat{\sigma}^+_{2j} \hat{\sigma}^-_{2j-1}+\text{H.c}+  h\hat{\sigma}_{2j}^x,
\end{align}
where $\Delta\gg  J,\lambda, h$.
To avoid the affection of the boundaries, without loss of generality, we choose periodic boundary condition, i.e., $ \sigma_{2L+1}= \sigma_1$.
Here, we use the Schrieffer-Wolff transformation~\cite{Schrieffer1966,Bravyi2011} to obtain the effective Hamiltonian
\begin{align}
\hat H_s = e^{-\hat S}\hat H e^{\hat S}.
\end{align}
To the second order, we have
\begin{align}  \label{hs} \nonumber
\hat H_s =  &\hat H_0 + (\hat H_I+[\hat H_0,\hat S])+\frac{1}{2}[ (\hat H_I+[\hat H_0,\hat S]),\hat S] \\
& + \frac{1}{2}[\hat H_I,\hat S].
\end{align}
Let $S = \sum_{n=1}^\infty \hat S^{(n)}$, where $\hat S^{(n)}$ is of order ($J^mh^{n-m}/\Delta^n$). Hence, $\hat H_s$ can be written as
\begin{align}  \label{hs_2}\nonumber
\hat H_s =  &\hat H_0 + \big(\hat H_I+[\hat H_0,\hat S^{(1)}]\big)+\big(\frac{1}{2} [\hat H_I,\hat S^{(1)}] \\
&+ [\hat H_0,\hat S^{(2)}]+\frac{1}{2}[\hat H_I+[\hat H_0,\hat S^{(1)}],\hat S^{(1)}] \big)+\dots.
\end{align}

\begin{figure*}[t]     		
	\includegraphics[width=0.7\textwidth]{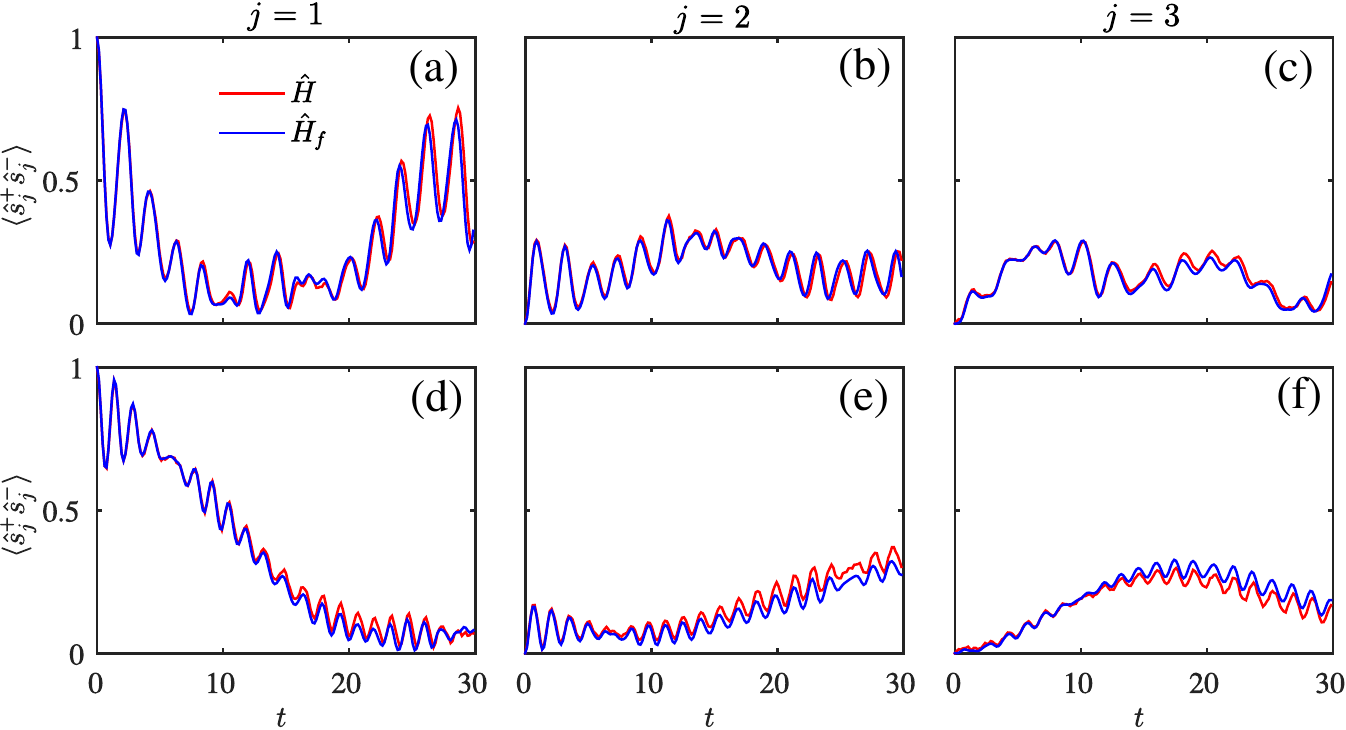}
	\caption{Quench dynamics of $\hat H$ (red lines) and $\hat H_f$ (blue lines) with $L=5$ under the periodic boundary condition. The initial state is $|\psi_0\rangle = |s\rangle\otimes|\tau\rangle$, of which $|s\rangle = \ket{\uparrow\downarrow\downarrow\downarrow\downarrow}$ and $|\tau\rangle=|\leftarrow\leftarrow\leftarrow\leftarrow\leftarrow\rangle$.
	We choose $J=20$, $\Delta=400$, thus $g_e=1$, $\lambda=2$.
	(a--c) $h =1$. (d--f) $h=2$.
	}
	\label{fig_s1}
\end{figure*}

Now we make an ansatz of transformation operator $\hat S$ as
\begin{align} \nonumber
&\hat S = \hat S^{(1)}+\hat S^{(2)},\\ \nonumber
&\hat S^{(1)} =  \sum_{j}^{L}A(\hat{\sigma}^+_{2j} \hat{\sigma}^-_{2j+1}+\hat{\sigma}^+_{2j} \hat{\sigma}^-_{2j-1}-\text{H.c}),\\
&\hat S^{(2)} =  \sum_{j}^{L}B(\hat{\sigma}^z_{2j} \hat{\sigma}^-_{2j+1}+\hat{\sigma}^z_{2j} \hat{\sigma}^-_{2j-1}-\text{H.c}),
\end{align}
Thus, we have
\begin{align}
&[\hat H_0,\hat S^{(1)}]= \sum_{j}^{L} -A(\Delta-\lambda) (\hat{\sigma}^+_{2j} \hat{\sigma}^-_{2j+1}+\hat{\sigma}^+_{2j} \hat{\sigma}^-_{2j-1}+\text{H.c}),\\ \nonumber
&[\hat H_0,\hat S^{(2)}]= \sum_{j}^{L} -B\Delta (\hat{\sigma}^z_{2j} \hat{\sigma}^-_{2j+1}+\hat{\sigma}^z_{2j} \hat{\sigma}^-_{2j-1}+\text{H.c}).
\end{align}
Let $A_j=J/(\Delta-\lambda)\simeq J/\Delta$, and we have
\begin{align} \nonumber
&\hat H_I+[\hat H_0,\hat S^{(1)}] =h\hat{\sigma}^x_{2j},\\ \nonumber
&\frac{1}{2}[\hat H_I,\hat S^{(1)}] =\sum_{j}^{L}
2JA(\hat{\sigma}^+_{2j+1} \hat{\sigma}^-_{2j+1}-\hat{\sigma}^+_{2j} \hat{\sigma}^-_{2j})\\ \nonumber
&\ \ \ \ +JA(\hat{\sigma}^+_{2j} \hat{\sigma}^z_{2j+1}\hat{\sigma}^-_{2j+2}-\hat{\sigma}^+_{2j-1} \hat{\sigma}^z_{2j}\hat{\sigma}^-_{2j+1}+\text{H.c})\\ \nonumber
&\ \ \ \ + \frac{1}{2}Ah (\hat{\sigma}^z_{2j} \hat{\sigma}^-_{2j+1}+\hat{\sigma}^z_{2j} \hat{\sigma}^-_{2j-1}+\text{H.c}),\\ \nonumber
&\frac{1}{2}[\hat H_I+[\hat H_0,\hat S^{(1)}],\hat S^{(1)}] \\
&\ \ \ \ \ \ \ = \frac{1}{2}Ah (\hat{\sigma}^z_{2j} \hat{\sigma}^-_{2j+1}+\hat{\sigma}^z_{2j} \hat{\sigma}^-_{2j-1}+\text{H.c}).
\end{align}
Let $B = Ah/\Delta$, we have
\begin{align}  \label{hs_3}\nonumber
\hat H_s = \sum_{j=1}^{L}& J^2/\Delta(\hat{\sigma}^+_{2j} \hat{\sigma}^z_{2j+1}\hat{\sigma}^-_{2j+2}-\hat{\sigma}^+_{2j-1} \hat{\sigma}^z_{2j}\hat{\sigma}^-_{2j+1}+\text{H.c})\\ \nonumber
+ & (\Delta+2J^2/\Delta)\hat{\sigma}^+_{2j-1}\hat{\sigma}^-_{2j-1} \\
+&(\lambda-2J^2/\Delta)\hat{\sigma}^+_{2j}\hat{\sigma}^-_{2j}
+  h\hat{\sigma}_{2j}^x.
\end{align}
Since the second term $\sum_{j=1}^{L}(\Delta+2J^2/\Delta)\hat{\sigma}^+_{2j-1}\hat{\sigma}^-_{2j-1}$ commutes the other terms, i.e., the system has spin $U(1)$ symmetry at the odd sites, this term can be neglected. In addition, we choose $\lambda=2J^2/\Delta$, so that the third term vanishes.
Labeling the odd and even qubits as $s_j$ and $\tau_{j+\frac{1}{2}}$, respectively, and $g_e = JA_j\simeq J^2/\Delta$,
we can obtain the final effective Hamiltonian of $\hat H$ as
\begin{align} \label{Hff} \nonumber
& \hat H_f =\hat H_1 + \hat H_2, \\ \nonumber
&\hat  H_1 =\sum_{j}^{L-1}g_e(\hat{s}^+_{j} \hat{\tau}^z_{j+\frac{1}{2}}\hat{s}^-_{j+1} +
\text{H.c.} ) +\sum_{j=1}^{L}h\hat{\tau}_{j+\frac{1}{2}}^x ,\\
& \hat H_2 = -\sum_{j}^{L-1}g_e(\hat{\tau}^+_{j+\frac{1}{2}} \hat{s}^z_{j+1}\hat{\tau}^-_{j+1+\frac{1}{2}}+ \text{H.c.}).
\end{align}

To verify the validity of the above derivation, we present numerical results in Fig.~\ref{fig_s1}.
Here, we numerically simulate the quench dynamics of the system by both $\hat H$ and $\hat H_f$.
According to Fig.~\ref{fig_s1}, we can find that the dynamics of $\hat H$ and $\hat H_f$ are almost the same.
Thus, $\hat H_f$ can indeed describe $\hat H$ effectively.

\begin{figure}[tb]     		
	\includegraphics[width=0.48\textwidth]{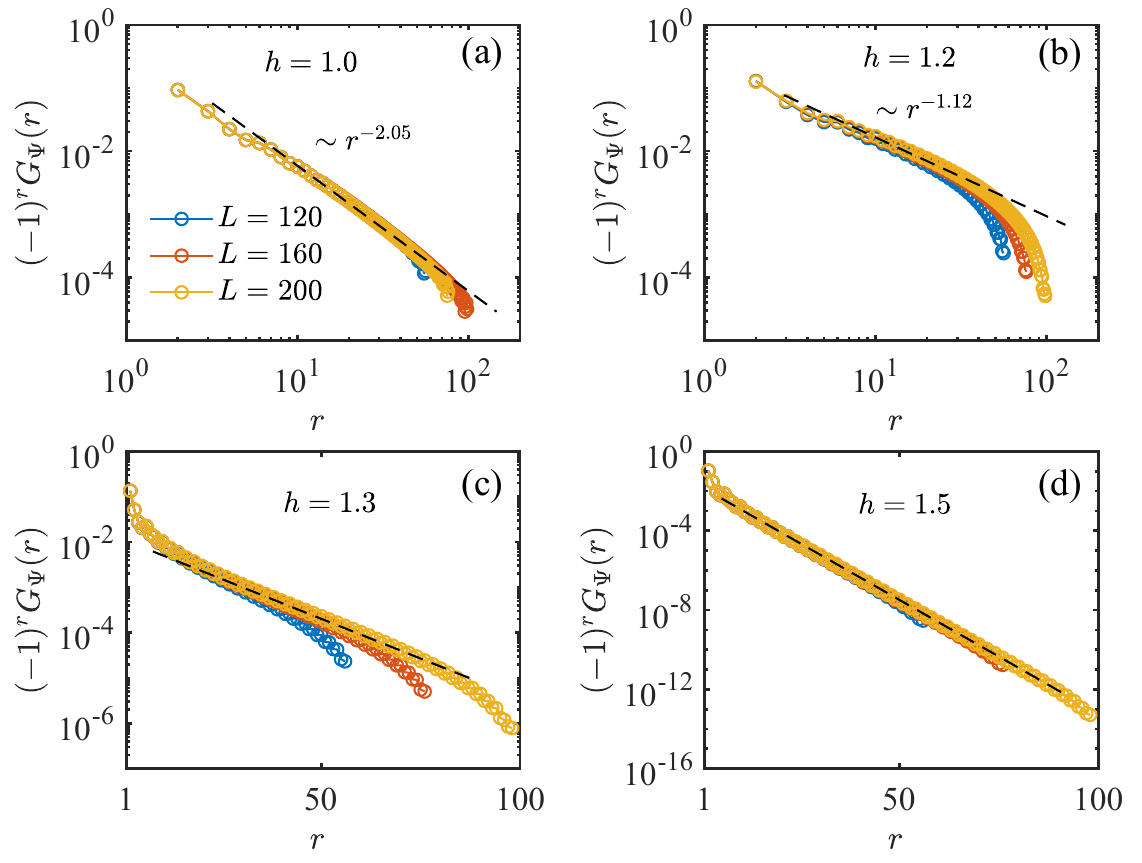}
	\caption{Correlation function $G_\Psi(r)=G_\Psi(L/2,L/2-r)$ for (a) $h=1.0$, (b) $h=1.2$, (c) $h=1.3$, and (d) $h=1.5$. The black dashed lines are linear fittings.
	}
	\label{fig_s3}
\end{figure}

\begin{figure}[b]     		
	\includegraphics[width=0.47\textwidth]{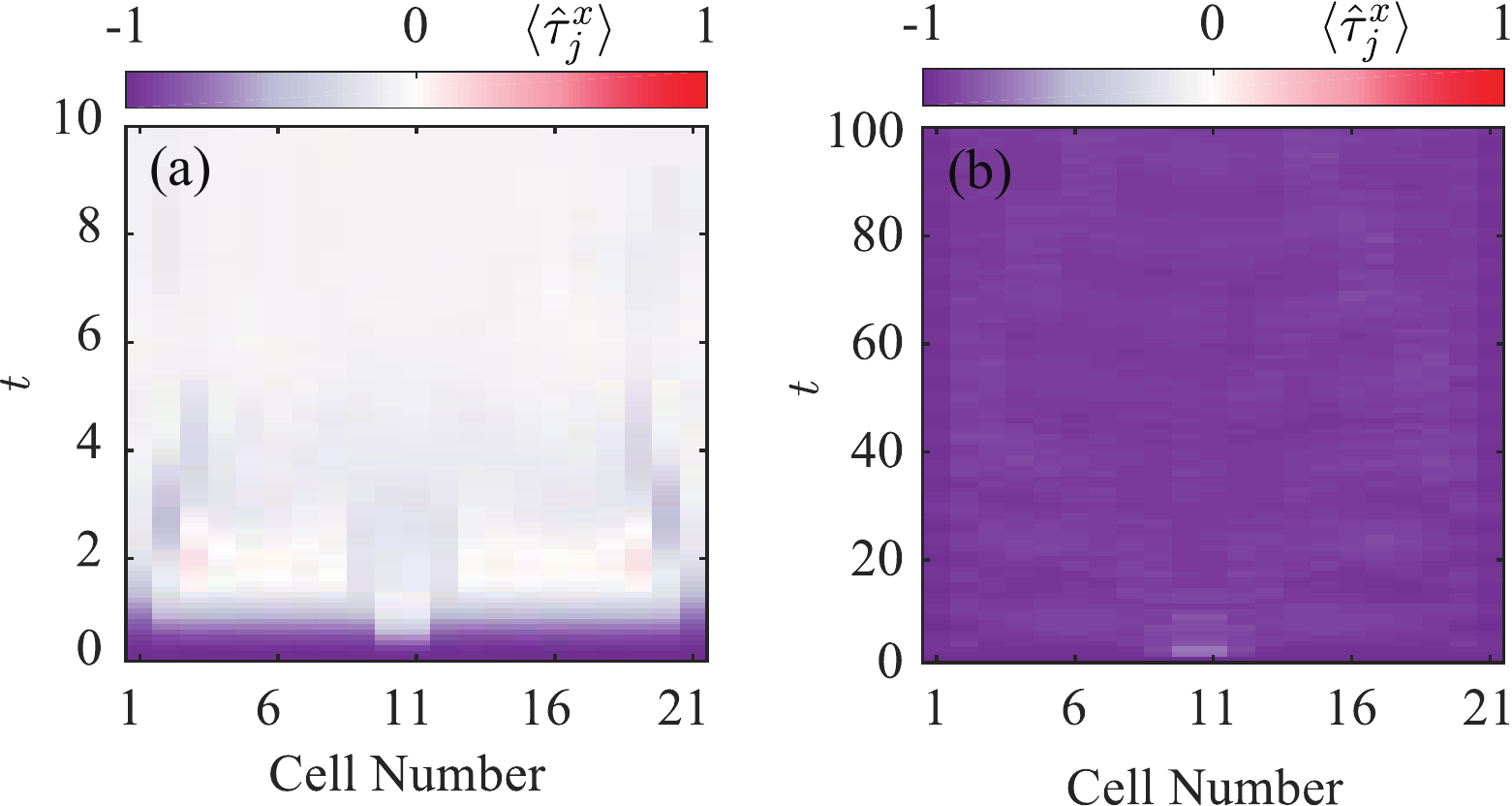}
	\caption{Time evolution of  $\braket{\hat\tau_j^x}:=\bra{\psi(t)} \hat{\tau}^x_j\ket{\psi(t)} $ for (a) $h=1$ and (b) $h=2$. The initial state is the same as one in Fig. 4 of main text. }
	\label{fig_s4}
\end{figure}

\section{Numerical Methods and Calculation Setup}\label{App2}
In this section, we introduce the detail of the numerical simulation. We use matrix product state (MPS) based methods to study the ground state and dynamical properties. A MPS is a kind of generic quantum many-body wave functions and has the form of matrices multiplication~\cite{Schollwock2005,Schollwock2011,Vidal2004}. It has been shown MPS can accurately describe the low energy states in one dimension and serves as the underlying wave function of density matrix renormalization group (DMRG) method~\cite{Schollwock2005,Schollwock2011}. We first use DMRG to obtain the ground state and study the static properties. For the quench dynamics, we apply the evolution operators onto the initial state and use the time evolution block decimation (TEBD) method to study the quench dynamics~\cite{Schollwock2005,Schollwock2011,Vidal2004}. The corresponding calculations are designed with TeNPy Library~\cite{Hauschild2018}.

\begin{figure*}[tb]     		
	\includegraphics[width=0.8\textwidth]{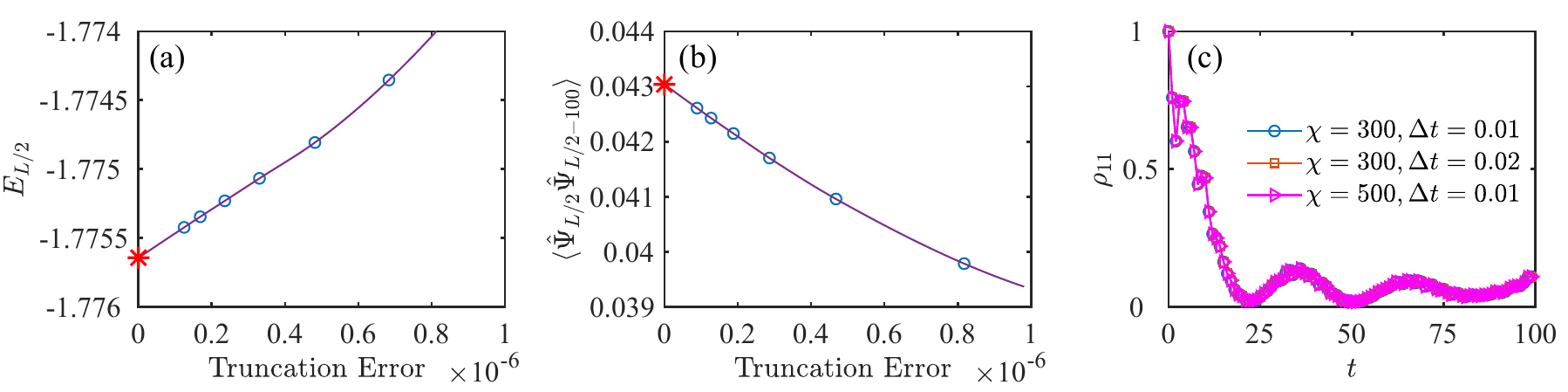}
	\caption{ Convergence for the physical quantities in the ground state and quench dynamics. (a) Finite truncation error analysis for the energy density at the central bond, in which $L=280$ and $h=1.245$. The blue circles are results calculated by DMRG methods, while the orange line is obtained by the extrapolation methods. (b) Finite truncation error analysis for the correlation function, in which $L=280$ and $h=1.245$. (c) Convergence of $\rho_{11}$ for $h=1.0$ with different $\chi$ and $\Delta t$. The blue circles denotes DMRG results. Here, $\chi$ is the maximum bond dimension and $\Delta t$ is the single time step of Suzuki-Trotte decomposition.}
	\label{fig_s2}
\end{figure*}

In the case of calculating the ground state, we adopt careful finite truncation error analysis to obtain the accurate result. By enlarging the maximum bond dimensions, we obtain more accurate result with smaller truncation errors. As shown in Figs.~\ref{fig_s2}(a--b), one can find both the local physical quantities and the correlations converge in the zero truncation error limit accurately.

For the quench dynamics, we also enlarge the maximum bond dimension to reduce the finite entanglement effect. Meanwhile we adopt smaller single time steps during the real time evolution to reduce the Suzuki-Trotte decomposition error. So that we obtain the converged result. In Fig.~\ref{fig_s2}(c), as an example, one can find the result has been converged when $\Delta t=0.02$ and $\chi=300$.

\section{Correlation Functions}\label{App3}
Here we focus on the correlations away from the critical point. In the weak field region $h<h_c$, the ground state is a gapless phase. The dimension of scaling operators is controlled by the Luttinger parameter. With different $h$, the correlation function should display power law behaviors with a varying exponent. As shown in Fig.~\ref{fig_s3} (a) and (b), $G_\Psi(i,j)$ indeed decay in a power law form and hosts different exponents for different $h$.

For the strong field region $h>h_c$, the ground state breaks the translational invariance and there is an energy gap. The correlation functions should decay exponentially with respect to the distance. As expected, Fig.~\ref{fig_s3} (c) and (d) verify the exponential form of $G_\Psi(i,j)$. Furthermore, for a larger $h$ the correlations decay much faster, which indicates the correlation length becomes shorter when enlarging $h$.

\section{Charge Spreading}\label{App4}
In the main text, we have discussed that the $s$ charge can display two different types of spreadings after a quantum quench.
It is diffusive under the weak transverse field, while in the case of strong transverse field, it is ballistic accompanied by the strong suppression of nearest-neighbor hopping.
We also analyze that this phenomenon is related to the confinement induced by the emergent Gauss law in the ground state.
In this section, we present a phenomenological  analysis to  interpret why $s$ charge is diffusive for small $h$ while it is ballistic for large $h$.

Under the weak transverse field, the $\tau$ sector can be hardly polarized at the $\hat\tau^x$ channel.
Thus, after a quick relaxation, the $\tau$ sector becomes a nearly random state, see Fig.~\ref{fig_s4}(a).
Therefore,  $\tau$ sector can be considered as a random scatting potential of $s$ sector leading to a diffusive spreading of $s$ charge.
In addition, one can find that the system in this case is a typical thermalized system, and the diffusion of $s$ charge can be considered as a reflection of thermalization.

In the case of strong transverse field, the system locates low-energy regime during the quench dynamics, so that the system fails to thermalization,
and $\tau$ sector is almost polarized in the $\hat\tau^x$ channel, see Fig.~\ref{fig_s4}(b).
Thus, there is too weak effective random potential to make $s$ charge spread diffusively.
In other words, the background of $s$ charge is so clean, that  it can transport ballistic.

On the other hand, this low-energy system is translational-symmetry broken leading to
an energy difference between the even and odd bonds, so that the hopping of $s$  charge between the nearest-neighbor
cells will violate the energy conservation. Therefore, despite that two nearest-neighbor s sites are resonant, i.e., the
corresponding two qubits have the same frequency, $s$ charge cannot hop to the nearest-neighbor cells. Instead, the
next-nearest-neighbor quantum tunneling is allowed, since the energy conservation is not violated in this case. From
this picture, we can know that the propagation of $s$  charge can indeed be a dynamical probe of spontaneous breaking
of translational symmetry.


	%
	

\begin{thebibliography}{64}%
		\makeatletter
		\providecommand \@ifxundefined [1]{%
			\@ifx{#1\undefined}
		}%
		\providecommand \@ifnum [1]{%
			\ifnum #1\expandafter \@firstoftwo
			\else \expandafter \@secondoftwo
			\fi
		}%
		\providecommand \@ifx [1]{%
			\ifx #1\expandafter \@firstoftwo
			\else \expandafter \@secondoftwo
			\fi
		}%
		\providecommand \natexlab [1]{#1}%
		\providecommand \enquote  [1]{``#1''}%
		\providecommand \bibnamefont  [1]{#1}%
		\providecommand \bibfnamefont [1]{#1}%
		\providecommand \citenamefont [1]{#1}%
		\providecommand \href@noop [0]{\@secondoftwo}%
		\providecommand \href [0]{\begingroup \@sanitize@url \@href}%
		\providecommand \@href[1]{\@@startlink{#1}\@@href}%
		\providecommand \@@href[1]{\endgroup#1\@@endlink}%
		\providecommand \@sanitize@url [0]{\catcode `\\12\catcode `\$12\catcode
			`\&12\catcode `\#12\catcode `\^12\catcode `\_12\catcode `\%12\relax}%
		\providecommand \@@startlink[1]{}%
		\providecommand \@@endlink[0]{}%
		\providecommand \url  [0]{\begingroup\@sanitize@url \@url }%
		\providecommand \@url [1]{\endgroup\@href {#1}{\urlprefix }}%
		\providecommand \urlprefix  [0]{URL }%
		\providecommand \Eprint [0]{\href }%
		\providecommand \doibase [0]{http://dx.doi.org/}%
		\providecommand \selectlanguage [0]{\@gobble}%
		\providecommand \bibinfo  [0]{\@secondoftwo}%
		\providecommand \bibfield  [0]{\@secondoftwo}%
		\providecommand \translation [1]{[#1]}%
		\providecommand \BibitemOpen [0]{}%
		\providecommand \bibitemStop [0]{}%
		\providecommand \bibitemNoStop [0]{.\EOS\space}%
		\providecommand \EOS [0]{\spacefactor3000\relax}%
		\providecommand \BibitemShut  [1]{\csname bibitem#1\endcsname}%
		\let\auto@bib@innerbib\@empty
	
		\bibitem{Buluta2009}I. Buluta and F. Nori,
		Quantum simulators,
		\href{http://science.sciencemag.org/content/326/5949/108}{Science \textbf{326}, 108 (2009)}.
		
		\bibitem{Georgescu2014}I. M. Georgescu, S. Ashhab, and F. Nori,
		Quantum simulation,
		\href{http://journals.aps.org/rmp/abstract/10.1103/RevModPhys.86.153}{Rev. Mod. Phys. \textbf{86}, 153 (2014)}.	


		\bibitem{Polkovnikov2011}A. Polkovnikov, K. Sengupta, A. Silva, and M. Vengalattore,
		Colloquium: Nonequilibrium dynamics of closed interacting quantum systems,
		\href{https://link.aps.org/doi/10.1103/RevModPhys.83.863}{Rev. Mod. Phys. \textbf{83}, 863 (2011)}
		
		\bibitem{Eisert2015}J. Eisert, M. Friesdorf, and C. Gogolin,
		Quantum many-body systems out of equilibrium,
		\href{https://www.nature.com/articles/nphys3215}{Nat. Phys. \textbf{11}, 124 (2015)}.


		\bibitem{Makhlin2001}Y. Makhlin, G. Sch\"{o}n, and A. Shnirman,
		Quantum-state engineering with Josephson-junction devices,
		\href{http://journals.aps.org/rmp/abstract/10.1103/RevModPhys.73.357}{Rev. Mod. Phys. \textbf{73}, 357 (2001)}.
		
		\bibitem{Gu2017}X. Gu, A. F. Kockum, A. Miranowicz, Y. Liu, and F. Nori,
		Microwave photonics with superconducting quantum circuits,
		\href{https://www.sciencedirect.com/science/article/pii/S0370157317303290}{Phys. Rep. \textbf{718}, 1 (2017)}.



			\bibitem{Salathe2015}Y. Salath\'{e}, M. Mondal, M. Oppliger, J. Heinsoo, P. Kurpiers, A. Poto\v{c}nik, A. Mezzacapo, U. Las Heras, L. Lamata, E. Solano, S. Filipp, and A. Wallraff,
			Digital quantum simulation of spin models with circuit quantum electrodynamics,
			\href{https://journals.aps.org/prx/abstract/10.1103/PhysRevX.5.021027}{Phys. Rev. X  \textbf{5}, 021027 (2015)}.
			
			
			\bibitem{Barends2015}R. Barends, L. Lamata, J. Kelly, L. Garc\'{\i}a-\'{A}lvarez, A. G. Fowler, A Megrant, E Jeffrey, T. C. White, D. Sank, J. Y. Mutus, B. Campbell, Yu Chen, Z. Chen, B. Chiaro, A. Dunsworth, I.-C. Hoi, C. Neill, P. J. J. O'Malley, C. Quintana, P. Roushan, A. Vainsencher, J. Wenner, E. Solano, and John M. Martinis,
			Digital quantum simulation of fermionic models with a superconducting circuit,
			\href{https://www.nature.com/articles/ncomms8654}{Nat. Commun.  \textbf{6}, 7654 (2015)}.


			\bibitem{Flurin2017}E. Flurin, V. V. Ramasesh, S. Hacohen-Gourgy, L. S. Martin, N. Y. Yao, and I. Siddiqi
			Observing topological invariants using quantum walks in superconducting circuits,
			\href{https://journals.aps.org/prx/abstract/10.1103/PhysRevX.7.031023}{ Phys. Rev. X \textbf{7}, 031023 (2017)}.


			\bibitem{Zhong}Y. P. Zhong,  D. Xu, P. Wang, C. Song, Q. J. Guo, W. X. Liu, K. Xu, B. X. Xia, C.-Y. Lu,
			S. Han, J.-W. Pan, and H. Wang,
			Emulating anyonic fractional statistical behavior in a superconducting quantum circuit,
			\href{https://journals.aps.org/prl/abstract/10.1103/PhysRevLett.117.110501}{Phys. Rev. Lett.  \textbf{117}, 110501 (2016)}.
			
			\bibitem{Roushan2017}P. Roushan, C. Neill, J. Tangpanitanon, V. M. Bastidas,
			A. Megrant, R. Barends, Y. Chen, Z. Chen, B. Chiaro,
			A. Dunsworth, A. Fowler, B. Foxen, M. Giustina, E. Jeffrey,
			J. Kelly, E. Lucero, J. Mutus, M. Neeley, C. Quintana, D. Sank,
			A. Vainsencher, J. Wenner, T. White, H. Neven, D. G. Angelakis, and J. Martinis,
			Spectroscopic signatures of localization with interacting photons in superconducting qubits,
			\href{http://science.sciencemag.org/content/358/6367/1175}{Science \textbf{358}, 1175 (2017)}.


           \bibitem{Xu2018}K. Xu, J. J. Chen, Y. Zeng, Y. R. Zhang, C. Song, W. X. Liu, Q. J. Guo, P. F. Zhang, D. Xu, H. Deng, K. Q. Huang,
			H. Wang, X. B. Zhu, D. N. Zheng, and H. Fan,
			Emulating many-body localization with a superconducting quantum processor,
			\href{https://journals.aps.org/prl/abstract/10.1103/PhysRevLett.120.050507}{Phys. Rev. Lett.  \textbf{120}, 050507 (2018)}.
			

						
			\bibitem{Song1}C. Song, D. Xu, P. Zhang, J. Wang, Q. Guo, W. Liu, K. Xu, H. Deng, K. Huang, D. Zheng, S.-B. Zheng, H. Wang, X. Zhu, C.-Y. Lu, and J.-W. Pan,
			Demonstration of topological robustness of anyonic braiding statistics with a superconducting quantum circuit,
			\href{https://journals.aps.org/prl/abstract/10.1103/PhysRevLett.121.030502}{Phys. Rev. Lett.  \textbf{121}, 030502 (2018)}.		
			

			
			\bibitem{Yan2019}Z. Yan, Y. R. Zhang, M. Gong, Y. Wu, Y. Zheng, S. Li, C. Wang,
			F. Liang, J. Lin, Y. Xu, C. Guo, L. Sun, C. Z. Peng, K. Xia, H. Deng, H. Rong, J. Q. You, F. Nori, H. Fan, X. Zhu, and J.-W. Pan,
			Strongly correlated quantum walks with a 12-qubit superconducting processor,
			\href{https://science.sciencemag.org/content/364/6442/753}{Science  \textbf{364}, 753 (2019)}.
			
			\bibitem{Ma2019}R. Ma, B. Saxberg, C. Owens, N. Leung, Y. Lu, J. Simon, and D. I. Schuster,
			A dissipatively stabilized Mott insulator of photons,
			\href{https://www.nature.com/articles/s41586-019-0897-9}{Nature  \textbf{566}, 51 (2019)}.
			
			\bibitem{Ye2019}Y. Ye, Z.-Y. Ge, Y. Wu, S. Wang, M. Gong, Y.-R. Zhang,
			Q. Zhu, R. Yang, S. Li, F. Liang, J. Lin, Y. Xu, C. Guo,
			L. Sun, C. Cheng, N. Ma, Z. Y. Meng, H. Deng, H. Rong,
			C.-Y. Lu, C.-Z. Peng, H. Fan, X. Zhu, and J.-W. Pan,
			Propagation and localization of collective excitations on a 24-Qubit superconducting processor,
			\href{https://link.aps.org/doi/10.1103/PhysRevLett.123.050502}{Phys. Rev. Lett. \textbf{123}, 050502 (2019)}.
			
			\bibitem{Guo2019}X.-Y Guo, C. Yang, Y. Zeng, Y. Peng, H.-K Li, H. Deng, Y.-R Jin, S. Chen, D.-N Zheng, and H. Fan,
			Observation of a dynamical quantum phase transition by a superconducting qubit simulation,
			\href{https://link.aps.org/doi/10.1103/PhysRevApplied.11.044080}{Phys. Rev. Applied \textbf{11}, 044080 (2019)}.		
			
			\bibitem{Xu2020}K. Xu, Z.-H Sun, W. Liu, Y.-R Zhang, H. Li, H. Dong, W. Ren, P. Zhang, F. Nori, D. Zheng, H. Fan, H. Wang,
			Probing dynamical phase transitions with a superconducting quantum simulator,
			\href{https://advances.sciencemag.org/content/6/25/eaba4935.full}{Science Advances \textbf{6}, eaba4935 (2020).}.
		
		\bibitem{Arute2019}F. Arute, K. Arya, R. Babbush, D. Bacon, J. C. Bardin, R. Barends, R. Biswas, S. Boixo, F. G.Brandao, D. A. Buell \textit{et al}.,
		Quantum supremacy using a programmable superconducting processor,
		\href{https://doi.org/10.1038/s41586-019-1666-5}{Nature (London) \textbf{574}, 505 (2019)}.		
		
		\bibitem{Zhong2020} H.-S. Zhong , H. Wang H, Y.-H. Deng,\textit{et al}.,
	  	Quantum computational advantage using photons,
		\href{https://www.science.org/doi/10.1126/science.abe8770}{2020 \textit{Science} \textbf{370} 1460}.
		
		\bibitem{Wu2021} Y. Wu, W.-S Bao, S. Cao,\textit{et al}.,
		Strong quantum computational advantage using a superconducting quantum processor
		\href{https://journals.aps.org/prl/abstract/10.1103/PhysRevLett.127.180501}{Phys. Rev. Lett. \textbf{127}, 180501(2021)}.
		
			
		
		\bibitem{Wilson1974}K. G. Wilson,
		Confinement of quarks,
		\href{https://link.aps.org/doi/10.1103/PhysRevD.10.2445}{Phys. Rev. D \textbf{10}, 2445 (1974)}.
		
		\bibitem{Kogut1979}J. B. Kogut,
		An introduction to lattice gauge theory and spin systems,
		\href{https://journals.aps.org/rmp/abstract/10.1103/RevModPhys.51.659}{Rev. Mod. Phys. \textbf{51}, 659 (1979)}.
		
		\bibitem{Wen2004}X. Wen, \textit{Quantum Field Theory of Many-Body Systems},
		Oxford Graduate Texts (Oxford University Press, Oxford,	2004)
		
		\bibitem{Fradkin2013}E. Fradkin, \textit{Field Theories of Condensed Matter Physics}
		(Cambridge University Press, Cambridge, England, 2013)
		
		\bibitem{Kitaev2003_1}A. Kitaev,
		Fault-tolerant quantum computation by anyons,
		\href{https://linkinghub.elsevier.com/retrieve/pii/S0003491602000180}{Ann. Phys. (Amsterdam) \textbf{303}, 2 (2003)}.		
		
		\bibitem{Kitaev2003_2}A. Kitaev,
		 Anyons in an exactly solved model and beyond,
		\href{https://linkinghub.elsevier.com/retrieve/pii/S0003491605002381}{Ann. Phys. (Amsterdam) \textbf{321}, 2 (2003)}.
		
		\bibitem{Zhou2016}Y. Zhou, K. Kanoda, and T.-K Ng,
		Quantum spin liquid states,
		\href{https://journals.aps.org/rmp/abstract/10.1103/RevModPhys.89.025003}{Rev. Mod. Phys. \textbf{89}, 025003 (2017)}.
		
      \bibitem{Senthil2004}
       T. Senthil, A. Vishwanath, L. Balents, S. Sachdev, and M. P. A. Fisher,
        Deconfined quantum critical points
       \href{https://science.sciencemag.org/content/303/5663/1490}{Science \textbf{303}, 1490 (2004)}.	
		
		\bibitem{Hebenstreit2013}F. Hebenstreit, J. Berges, and D. Gelfand,
		Real-time dynamics of string breaking,
		\href{https://link.aps.org/doi/10.1103/PhysRevLett.111.201601}{Phys. Rev. Lett. \textbf{111}, 201601 (2013)}.
		
		\bibitem{Kormos2017}M. Kormos, M. Collura, G. Tak\'{a}cs, and P. Calabrese,
		Real-time confinement following a quantum quench to a non-integrable model,
		\href{https://www.nature.com/articles/nphys3934}{Nat. Phys. 13, \textbf{246} (2017)}.
		
		
		
		
		\bibitem{Zohar2012}E. Zohar, J. I. Cirac, and B. Reznik,
		Simulating compact quantum electrodynamics with ultracold atoms:
		probing confinement and nonperturbative effects,
		\href{https://link.aps.org/doi/10.1103/PhysRevLett.109.125302}{Phys. Rev. Lett. \textbf{109}, 125302 (2012)}.
		
		
		\bibitem{Banerjee2012}D. Banerjee, M. Dalmonte, M. M\"{a}uller, E. Rico, P. Stebler, U.-J. Wiese, and P. Zoller,
		Atomic quantum simulation of dynamical gauge fields coupled to fermionic matter:
		from string breaking to evolution after a quench,	
		\href{https://link.aps.org/doi/10.1103/PhysRevLett.109.175302}{Phys. Rev. Lett. \textbf{109}, 175302 (2012)}.
		
		\bibitem{Barbiero2019}L. Barbiero, C. Schweizer, M. Aidelsburger, E. Demler, N. Goldman, F. Grusdt,
		Coupling ultracold matter to dynamical gauge fields in optical lattices: From flux attachment to $\mathbb{Z}_2$ lattice gauge theories
		\href{https://advances.sciencemag.org/content/5/10/eaav7444}{Sci. Adv. \textbf{5}, 7444 (2019)}.
		
		\bibitem{Hauke2017}P. Hauke, D. Marcos, M. Dalmonte, and P. Zoller,
		Quantum simulation of a lattice Schwinger model in a chain of trapped ions,			
		\href{https://link.aps.org/doi/10.1103/PhysRevX.3.041018}{Phys. Rev. Lett. \textbf{3}, 041018 (2013)}.	
		
		
	   \bibitem{Marcos2013}D. Marcos, P. Rabl, E. Rico, and P. Zoller,
	    Superconducting circuits for quantum simulation of dynamical gauge fields,				
	   \href{https://link.aps.org/doi/10.1103/PhysRevLett.111.110504}{Phys. Rev. Lett. \textbf{111}, 110504 (2013)}.
		
	   \bibitem{Brennen2016}G. K. Brennen, G. Pupillo, E. Rico, T. M. Stace, and D. Vodola,	
	   Loops and strings in a superconducting lattice gauge simulator	   			
       \href{https://link.aps.org/doi/10.1103/PhysRevLett.117.240504}{Phys. Rev. Lett. \textbf{117}, 240504 (2016)}.				
		
		\bibitem{Zohar2017}E. Zohar, A. Farace, B. Reznik, and J. I. Cirac,
		Digital quantum simulation of $\mathbb{Z}_2$ lattice gauge theories	with dynamical fermionic matter, 			
		\href{https://link.aps.org/doi/10.1103/PhysRevLett.118.070501}{Phys. Rev. Lett. \textbf{118}, 070501 (2017)}.
		
	    \bibitem{Chamon2020}C. Chamon, D. Green, and Z.-C Yang,
	    Constructing quantum spin liquids using combinatorial gauge symmetry, 			
	    \href{https://link.aps.org/doi/10.1103/PhysRevLett.125.067203}{Phys. Rev. Lett. \textbf{125}, 067203 (2020)}.
	
	
	
				
		\bibitem{Schweizer2019}C. Schweizer, F. Grusdt, M. Berngruber, L. Barbiero, E. Demler,
		N. Goldman, I. Bloch, and M. Aidelsburger,
		Floquet approach to $\mathbb{Z}_2$ lattice gauge theories with
		ultracold atoms in optical lattices,
		\href{https://www.nature.com/articles/s41567-019-0649-7}{Nat. Phys. \textbf{15}, 1168 (2019)}.
		
		
		\bibitem{Gorg2019}F. G\"{o}rg, K. Sandholzer, J. Minguzzi, R. Desbuquois, M. Messer, and
		T. Esslinger,
		Realization of density-dependent Peierls phases
		to engineer quantized gauge fields coupled to
		ultracold matter
		\href{https://www.nature.com/articles/s41567-019-0615-4}{Nat. Phys. \textbf{15}, 1161 (2019)}.
		

	
		
		\bibitem{Yang2019}B. Yang, H. Sun, R. Ott, H.-Y Wang, T. V. Zache, J. C. Halimeh, Z.-S. Yuan, P. Hauke, and J.-W. Pan,
		Observation of gauge invariance in a 71-site quantum simulator,
		\href{https://arxiv.org/abs/2003.08945}{arXiv:2003.08945}.
			

		\bibitem{Schrieffer1966}J. R. Schrieffer, and P. A.Wolff,
        Relation between the Anderson and Kondo Hamiltonians, 			
       \href{https://journals.aps.org/pr/abstract/10.1103/PhysRev.149.491}{Phys. Rev. \textbf{149}, 491 (2020)}.		

        \bibitem{Bravyi2011}S. Bravyi, D. P. DiVincenzo, and D. Loss,
         Schrieffer-Wolff transformation for quantum many-body systems, 			
         \href{https://doi.org/10.1016/j.aop.2011.06.004}{Ann. Phys. (Amsterdam) \textbf{326}, 2793 (2011)}.
		
		
		\bibitem{Borla2020}U. Borla, R. Verresen, F. Grusdt, and S. Moroz,
		Confined phases of one-dimensional spinless fermions coupled to $\mathbb{Z}_2$ gauge theory, 			
		\href{https://link.aps.org/doi/10.1103/PhysRevLett.124.120503}{Phys. Rev. Lett. \textbf{124}, 120503 (2020)}.
		

		\bibitem{Schollwock2005}U. Schollw\"{o}ck,
         The density-matrix renormalization group,
        \href{https://link.aps.org/doi/10.1103/RevModPhys.77.259}{Rev. Mod. Phys. \textbf{77}, 259 (2005)}.

        \bibitem{Schollwock2011}U. Schollw\"{o}ck,
        The density-matrix renormalization group in the age of matrix product states,
        \href{https://doi.org/10.1016/j.aop.2010.09.012}{Ann. Phys. \textbf{326}, 96 (2011)}.

        \bibitem{noteGL}
        Here, $f(i,j)=\frac{(j-i+1)(j+i+2)}{2}$ is an integer.
        The sign of charge $\hat W_{s_z}$ is consistence with the polarization of $\hat s_z$.

        \bibitem{cite_op_dim}
        $\hat{\Psi}_i$ has the same dimension as $\hat{\Psi}_{2j}-\hat{\Psi}_{2j-1}$ .

        \bibitem{Vidal2004}G. Vidal,
        Efficient simulation of one-dimensional quantum many-body systems,
        \href{https://link.aps.org/doi/10.1103/PhysRevLett.93.040502}{Phys. Rev. Lett. \textbf{93}, 040502 (2004)}.

        \bibitem{Kim2013}H. Kim, and D. A. Huse,
         Ballistic spreading of entanglement in a diffusive nonintegrable system
         \href{https://link.aps.org/doi/10.1103/PhysRevLett.111.127205}{Phys. Rev. Lett. \textbf{11}, 127205 (2013)}.
		

         \bibitem{Hauschild2018}J. Hauschild, and F. Pollmann,
         Efficient numerical simulations with Tensor Networks: Tensor Network Python (TeNPy),
        \href{https://scipost.org/10.21468/SciPostPhysLectNotes.5}{SciPost Phys. Lect. Notes, 5 (2018)}.



	\end{thebibliography}
\end{document}